\documentclass[12pt]{article}
\usepackage[dvips]{graphicx}
\usepackage{subfigure}
\usepackage{color}
\usepackage{mathrsfs}
\usepackage{amsfonts,amssymb,amsmath}
\usepackage{bm}
\oddsidemargin=\evensidemargin

\allowdisplaybreaks

\def\ra{\rightarrow}

\def\p{\mathbb P}
\def\e{\mathrm{e}}
\def\veps{\varepsilon}

\numberwithin{equation}{section}

\begin{document}
\title{Estimation and prediction of credit risk based on rating transition systems\footnote{Supported in
 part by the National Natural Science Foundation of China (Grant No. 11771327, 11431014)}}
\author{Jinghai Shao${}^a$\footnote{Email: shaojh@tju.edu.cn},\quad Siming Li${}^b$,\quad Yong Li${}^b$\\[0.15cm]
{\small a: Center for Applied Mathematics, Tianjin University, Tianjin 300072, China}\\
{\small b: School of Statistics, Beijing Normal University, Beijing 100875, China}
}
\date{}
\maketitle

\begin{abstract}
Risk management is an important practice in the banking industry. In this paper we develop a new methodology to estimate and predict the probability of default based on the rating transition matrices, which relates the rating transition matrices directly to the macroeconomic variables without using a latent variable. Our method extends the framework of one factor model (Belkin et al. 1998). Especially, it is useful in predicting the PD and doing stress testing. Simulation is conducted to demonstrate the effectiveness of this new method compared with one factor model.
\end{abstract}

\section{Introduction}

Credit risk management has received tremendous attention from the bank industry. Various methods were proposed concerning the estimation and prediction of key risk parameters like probability of default (\textbf{PD}). A large volume of literature on credit risk management has evolved. This development was partly reflected by supervisors under the framework of Basel II and Basel III.

In their seminal work on credit spread, Jarrow et al. \cite{Jar} derived the risk premium for the credit risk process from a Markov chain (discrete time or continuous time) on a finite state space. Then the estimate of \textbf{PD} is derived from the transition matrix of the Markov chain. Then, the works such as Wilson \cite{Wilson}, Belkin et al.  \cite{Bel-a,Bel-b}, Alessandrini  \cite{Ale}, Kim \cite{Kim}, Nickell et al.  \cite{Nick} identify the important impact of the business cycle on the rating transition matrix. Belkin et al.\! \cite{Bel-b} established a one-factor model to measure the business cycle and proposed a method of calculating  rating transition matrices conditional on the credit risk. Nickell et al. \cite{Nick} quantified the dependence of rating transition matrices on the industry, domicile of the obligor, and the stage of the business cycle by employing ordered probit model.
Since one-factor model rules out the possibility that different ratings respond to the same credit condition change at different rates,  Wei \cite{Wei} established a multi-factor model to overcome this shortcoming and meanwhile retained the main feature of one-factor model \cite{Bel-b}. As the models given by \cite{Bel-b} and \cite{Wei} provide a method to obtain fitted rating transition matrices from the variables measured the business cycle, these models can be used in the prediction of rating transition matrix, specifically \textbf{PD}, once one get a reasonable prediction of the business cycle. Particularly, this is useful in credit portfolio stress testing.  For their simplicity and efficiency,   one-factor model \cite{Bel-b} or multi-factor model \cite{Wei} are still practices in the banking industry till today.

However, despite the development of Wei  \cite{Wei}, there are also some shortcomings in the one-factor model proposed by Belkin et al.\! \cite{Bel-b}. We shall introduce these shortcomings and provide a new model to overcome them in current work.
In order to make our statements clear, we need to first recall the main idea of \cite{Bel-b}.

According to \cite{Bel-b},  a underlying random variable $X$ is used to reflect rating migrations, and $X$ is assumed to be {\it normally distributed}. Conditional on an initial credit rating grade $i$ at the beginning of a year, partition the value domain of $X$ into a set of disjoint intervals $(x_{j-1}^i,x_j^i)$ and  $i,j$ belong to the set $\{1,\ldots,K\}$, which stands for the set of rating grades.
These series of thresholds satisfy $-\infty=x_0^i<x_1^i<\ldots<x_{K-1}^i<x_K^i=+\infty$. The random variable $X$ measures the changes in creditworthiness, and these changes are supposed to be caused by idiosyncratic variation and systematic variation. Hence, $X$ is assumed to be split into two parts as follows:
\begin{equation}
  \label{e-1} X=\sqrt{1-\rho}\,Y+\sqrt{\rho} Z,
\end{equation}
where $Y$ represents idiosyncratic component, unique to a borrower, $Z$ stands for a systematic component, shared by all borrowers, and $\rho$ is a constant taking values in the interval $[0,1]$. $Z$ is used to measure the credit cycle, and is independent of $Y$.  In order to ensure that $X$ given by \eqref{e-1} is still normally distributed, it is assumed that $Y$ and $Z$ are all standard normally distributed variables. Hence, $\rho$ is the correlation parameter between $X$ and $Z$, which represents the percentage of the variance of $X$ explained by $Z$. Suppose that a borrower is initially at the rating grade $i$. If its creditworthiness $X$ is located at the interval $(x_{j-1}^i,x_j^i]$, then this borrower will be at the grade $j$ this year. Let $P_{ij,t}$ be the transition probability from rating grade $i$ to rating grade $j$ at time $t$ conditional on the given systematic component $Z_t=z$.  The standard normal distribution of $Y_t$ can easily yield that
\begin{equation}\label{m-factor}
P_{ij,t}(z)=\Phi\Big(\frac{x_{j}^i-\sqrt{\rho}z}{\sqrt{1-\rho}}\Big)
-\Phi\Big(\frac{x_{j-1}^i-\sqrt{\rho}z}{\sqrt{1-\rho}}\Big),\ \ i,j\in\{1,\ldots,K\},
\end{equation}
where $\Phi(x)$ denotes the cumulative distribution function of the standard normal variable. According to the law of large number, direct calculation under the help of the assumption on the distribution of $Z$ can yield that
\[\frac1{N}\sum_{t=1}^N P_{ij,t}\quad \text{converges almost surely to}\ \ \Phi(x_j^i)-\Phi(x_{j-1}^i) \ \text{as}\ N\ra +\infty.\]
So it is reasonable to use the average rating transition probability $\bar P_{ij}=\frac 1N\sum_{t=1}^NP_{ij,t}$ to provide an estimation $\{\hat{x}_j^i;\,j=1,\ldots,K\}$ of the thresholds $\{x^i_j;\,j=1,\ldots, K\}$. Precisely,
\begin{equation}\label{tres}
\hat x_j^i=\Phi^{-1}\Big(\sum_{k=1}^j \bar P_{ik}\Big),\ \ \text{and}\ \hat{x}_j^i\longrightarrow x_j^i\ a.s. \ \text{as}\ N\rightarrow \infty.
\end{equation}
Replacing $x_j^i$ by $\hat{x}_j^i$ in \eqref{m-factor} and given $z$, we obtain the fitted rating transition probability
\begin{equation}
  \label{est-prb} \hat P_{ij,t}(z)=\Phi\Big(\frac{\hat x_j^i-\sqrt{\rho} z}{\sqrt{1-\rho}}\Big)-\Phi\Big(\frac{\hat x_{j-1}^i-\sqrt{\rho} z}{\sqrt{1-\rho}}\Big),\ \ i,j\in\{1,\ldots,K\}.
\end{equation}
The least-squares problem takes the form: for fixed $\rho$ and $t$,
\begin{equation}
  \label{e-2}\min_{z}\sum_{i=1}^{K-1}\sum_{j=1}^{K} \frac{n_{t,i}\big(P_{ij,t}-\hat P_{ij,t}(z)\big)^2}{\hat P_{ij,t}(z)(1-\hat P_{ij,t}(z))},
\end{equation}
where $P_{ij,t}$ represents the $i$ to $j$ transition probability observed in time $t$ and $n_{t,i}$ the number of borrowers from initial grade $i$ observed in that time period. The solution of \eqref{e-2} provides a value $\hat{Z}_t$ for the systematic component for  every time $t$. The series of numbers $(\hat{Z}_t)_{t}$ is called {\em extracted $m$-factor} sometimes. Inserting the number $\hat{Z}_t$ into \eqref{est-prb}, we can also obtain an estimation of the rating transition probability
\begin{equation}\label{est-rate}
\hat{P}_{ij,t}=\hat{P}_{ij,t}(\hat{Z}_t)=\Phi\Big(\frac{\hat x_j^i-\sqrt{\rho} \hat{Z}_t}{\sqrt{1-\rho}}\Big)-\Phi\Big(\frac{\hat x_{j-1}^i-\sqrt{\rho} \hat{Z}_t}{\sqrt{1-\rho}}\Big).
\end{equation}
Formula \eqref{est-rate} is used to predict the transition probability matrices in practice when the predicted $Z_t$ is provided.

The previous procedure is the main step introduced by Belkin et al.\! \cite{Bel-b} to derive the fitted rating transition matrix from the representative variable $Z$ of the credit cycle. There are two kinds of shortcomings in this methodology. First, the assumption on the distributions of $X$ and further on $Y$ and $Z$ is not natural and hard to be verified. In the practical application, the extracted series of $(Z_t)$ tends to show that $Z$ is not standard normally distributed. This assumption also causes the
appearance of the
correlation coefficient $\rho$. This quantity $\rho$ is of important economic meaning, however, its value is very hard to be estimated.  Second, the data quantity to extract the series of $Z_t$ is very limited by the size of transition probability matrix according to \eqref{e-2}, which is up to $K\times (K-1)$. When one uses the multi-factor model in Wei \cite{Wei}, the data quantity in this step is even smaller.

To predict the probability of default and to do stress testing are the most important application of the methodology of Belkin et al.\! \cite{Bel-b}, which are widely used in practice.  As variable $Z$ is intended for characterizing the business cycle, one can establish a regression model to link it with some macroeconomic variables. Then the prediction of the macroeconomic variables will lead to a prediction of variable $Z$, and hence yield a prediction of rating transition matrix and \textbf{PD} by \eqref{est-prb}. For example, consider the following regression model in order to get the forecast \textbf{PD}:
\begin{equation}\label{reg}
Z=\beta_0+\beta_1 \xi_1+\beta_2 \xi_2+\veps,
\end{equation}
where $\xi_1$ denotes the GDP growth, $\xi_2$ denotes the capacity utilization rate.
Once we fix the coefficient $\beta_0$, $\beta_1,\,\beta_2$ using the extracted value $(\hat{Z}_t)_t$ of $Z$ by \eqref{e-2} and the observations $(\xi_{1,t})_t$ and $(\xi_{2,t})_t$ of $\xi_1,\,\xi_2$, we can obtain the forecast value of $Z$ by the forecast value of $\xi_1$ and $\xi_2$. Inserting this value into formula \eqref{est-rate}, we can get the forecast rating transition probability as desired. However, the norm distribution assumption of $Z$ inheriting from model \eqref{e-1} is very strange in the regression model. The important contribution of Belkin et al.\! \cite{Bel-b} is to connect rating transition probability to a latent variable $Z$ characterizing the business cycle.  However,  its assumption on the distributions of $X$ and $Z$ restricts further application of this model.

In this work we propose a new credit risk model, which is called for simplicity a \textit{macroeconomic-risk model}. Analogous to Belkin et al. \cite{Bel-b}, we also describe the change of rating grades from two aspects: idiosyncratic variation and systematic variation.  However, instead of using a latent variable to describe systematic variation, we use directly some macroeconomic variables as a representation of the business cycle. The important advantage of this model is that there is no priori assumption on the distributions of $X$ and $Z$.
To be more precise, we also split it into two parts: systematic part, idiosyncratic part. We only assume that the idiosyncratic part is of normal distribution. The main difference to \cite{Bel-b} lies in the systematic part. In \cite{Bel-b}, this part is assumed to be standard normally distributed, however, there is no observed values to check it. In our model we shall use some macroeconomic variables to reflect the business cycle, and hence there are observed values for this part and no priori assumption on the distributions of these macroeconomic variables are needed. In some sense our macroeconomic-risk model can be viewed as a kind of combination of \eqref{e-1} and \eqref{reg}. Analogous to \cite{Bel-b}, the transition probabilities are related to the business cycle based on the location of $X$ in its range. We then can use the transformation of rating transition matrices as dependent variable and the macroeconomic variables as explanatory variables to establish a multivariate linear regression model. This method can overcome the shortcomings of Belkin et al.\! \cite{Bel-b} mentioned above, and remains to be simple enough to be applied in practice. Moreover, it can improve the efficiency of \cite{Bel-b}, which is demonstrated by simulation in the last section.

This work is organized as follows. In Section 2, we give a more precise description on the methodology of \cite{Bel-b} and introduce its limitation. In Section 3, we establish the macroeconomic-risk model and introduce how to use it to forecast \textbf{PD} or do stress testing.
In Section 4, we do some simulation and comparison on the efficiency between the approach of \cite{Bel-b} and that of our macroeconomic-risk model in predicting \textbf{PD}.

\section{Overview of One Factor Model}

In this section, we give more details on the one factor model in \cite{Bel-b}. There are mainly three issues: (i) the assumption on the distributions of $X$ and $Z$; (ii) the data quantity when solving least square problem; (iii) further application of the extracted time series $(Z_t)_t$.

Throughout this work, we assume a rating system with $K$ rating grades, where the $K$-th grade is the default grade. A 1-year transition matrix is a $(K-1)\times K$ matrix with the probabilities that a debtor in rating grade $i$ migrates to rating grade $j$ within 1 year. Following Jarrow et al.\! \cite{Jar}, the rating transition matrix are modeled as the transition probability matrix of a Markov chain. There are mainly two methods to estimate the transition probability matrices, that is, the cohort method and the duration method. See, for instance, Engolmann and Ermakov \cite{EE}. According to the discussion in \cite{EE}, it is suitable to use the cohort method in our case to derive the empirical transition probability matrix.  Namely,
\begin{equation}
  \label{emp-matr} \tilde P_{ij,t}=\frac{N_{ij,t}}{N_{i,t}},
\end{equation}
where $N_{i,t}$ is the number of debtors in rating grade $i$ at the beginning of time period $t$, and $N_{ij,t}$ is the number of rating transitions from rating grade $i$ to grade $j$ during this period. In \cite{Bel-b}, to solve the least square problem \eqref{e-2},  $\tilde P_{ij,t}$ defined by \eqref{emp-matr} is indeed used to represent $P_{ij,t}$ in \eqref{e-2}.
Consequently, increasing the number of debtors in each rating grade can improve the approximation accuracy of rating transition matrices $(P_{ij,t})$ by empirical rating transition matrices $(\tilde P_{ij,t})$ due to the law of large numbers.

As mentioned in the introduction, due to the assumption on the distributions of $X$ and $Z$, the estimation of  $\rho$ in \eqref{e-1} is also a question. Belkin et al. suggested the following approach to estimate $\rho$. Apply  the minimization in \eqref{e-2}, using an assumed value of $\rho$. Then this yields a time series for $Z_t$ conditional on $\rho$. Repeat this procedure for many values of $\rho$ and use a numerical search procedure to find the particular $\rho$ value for which $Z_t$ time series has variance of one.
This method is not effective and it relies on the viewpoint that the time series $Z_t$ are reasonable sampling of the variable $Z$. On the other hand, if one use other method to estimate $\rho$,  for instance, the Basel III asset correlation formula, the obtained estimate $\rho$ can not guarantee the basic assumption on the distributions of $X$ and $Z$ of the one factor model in Belkin et al. \cite{Bel-b}.

It is also worthy noting that the data quantity in solving the least-squares problem \eqref{e-2} in \cite{Bel-b} is poor. All the data could be used to extract  $Z_t$ is at most $K\times (K-1)$. The data quantity could be improved by increasing the number of grades. However, it will decrease the effectiveness of $\tilde P_{ij,t}$ to approximate the rating transition matrix $P_{ij,t}$.

As a parameter which is created to measure the business cycle, $Z_t$ has no directly observations to back it. In application of this one factor model to forecast \textbf{PD} or to do stress testing, it is necessary to use extracted time series  $Z_t$ to run a predicting model. One needs to pay more attention on the risk in this process. Since there is no estimator who can perfectly estimate a parameter, one has to bear the risk caused by using an estimator set rather than an observation set that the predictor may very much likely magnify the input error.

\section{A Macroeconomic-risk model}
In this section we present mathematical deduction of the macroeconomic-risk model. We use a latent random variable $X$  to reflect rating migrations. Compared with Belkin et al. \cite{Bel-b}, we do not make any assumption directly on the distribution of $X$. We assume that $X$ can be split into two parts as follows:
\begin{equation}
  \label{e-3} X={\bm M^T\bm \beta} +Y,
\end{equation}
where $\bm M=(M_1,\ldots,M_n)^T$ is a vector of given macroeconomic variables  which is used to represent  business cycle, $Y$ represents idiosyncratic component which is assumed to follow a standard normal distribution. Here and in the sequel $A^T$ stands for the transpose of a matrix $A$. Assume that $Y,\,\bm{M}$ are mutually independent.

Suppose that a borrower is initially at the rating grade $i$ and there exists an underlying set of thresholds $-\infty=x_0^i<x_1^i<\ldots<x_{K-1}^i<x_{K}^i=+\infty$. To emphasize the initial rating grade, we write $X_t^i$ instead of $X_t$. If $X_t^i$ is located at the interval $(x_{j-1}^i,x_{j}^i]$,   then the borrower will be at the grade $j$ at the end of this time period. Consequently,
\begin{align*}
  P_{ij,t}(\bm m)&:=\p(X_t^i\in (x_{j-1}^i,x_{j}^i]|\bm M_t=\bm m,)\\
               &=\p(x_{j-1}^i<\bm m^T\bm \beta^i+Y<x_j^i)\\
               &=\Phi(x_{j-1}^i-\bm m^T\bm \beta^i)-\Phi(x_{j}^i- \bm m^T\bm \beta^i).
\end{align*}
For computational reasons, in the sequel we use right accumulative function \[\Phi(x)=\int^{+\infty}_x\frac{1}{\sqrt{2\pi}}\e^{-r^2/2}\mathrm{d} r.\]
\begin{equation}
\Phi(x_j^i-\bm m^T\bm \beta^i)=\sum_{k=j+1}^K P_{ik,t}(\bm m).
\end{equation}
We obtain that
\begin{equation}
  \label{e-4} \Phi^{-1}\Big(\sum_{k=j+1}^K P_{ik,t}(\bm m)\Big)=x_j^i-\bm m^T\bm \beta^i.
\end{equation}
Set $U_{ij,t}=\Phi^{-1}\Big(\sum_{k=j+1}^K P_{ik,t}(\bm m)\Big)$, then \eqref{e-4} can be  rewritten as
\begin{equation*}
U_{ij,t}=x_j^i-{\bm m}^T{\bm \beta}^i.
\end{equation*}
Applying the estimator $\tilde P_{ij,t}$ defined by \eqref{emp-matr}  to estimate $P_{ij,t}(\bm m,y)$, we establish a regression model
\begin{equation}\label{e-5}
\widetilde U_{ij,t}:=\Phi^{-1}\Big(\sum_{k=j+1}^K \tilde P_{ij,t}\Big)=x_j^i-\bm M_t^T\bm \beta^i+\veps_t,
\end{equation}
where $\veps_t$ is the error term caused by this replacement.
The equation \eqref{e-5} can be rewritten  in the form
\begin{equation}
  \label{e-6} \widetilde U_{ij,t}=(1,\bm M_t^T)\begin{pmatrix}
    x_j^i\\ -\bm \beta^i \end{pmatrix}  +\veps_t,
\end{equation}

As a multivariate regression model,
the estimation of $x_j^i$ and $\bm \beta^i$ can be given by least square estimators. Let $\hat x_j^i$ and $\hat{ \bm \beta^i}$ be the least square estimators of $x_j^i$ and $\bm \beta^i$ respectively. Due to \eqref{e-5}, we get the estimation $\hat U_{ij,t}$ of $\widetilde U_{ij,t}$, namely,
\begin{equation}\label{U}
\hat U_{ij,t}=\hat x_j^i-\hat{ \bm M_t}^T\hat{ \bm \beta^i}.
\end{equation}
Invoking the definition of $U_{ij,t}$, we finally obtain a fitted rating transition probability $\hat P_{ij,t}$ by
\begin{equation}\label{e-7}
\hat P_{ij,t}=\Phi(\hat U_{i(j-1),t})-\Phi(\hat U_{ij,t}),
\end{equation}
which is exactly the desired quantity.

The current model is easy to be used to forecast \textbf{PD} or to do stress testing. For these purpose, one only need to insert the prediction of macroeconomic variables $\bm M_t$ into \eqref{U} and \eqref{e-7} to obtain the desired results.


The macroeconomic-risk model has two main advantages from theoretical point of view: (i) No assumption on priori distribution of latent variable $Z$ which is used to measure the business cycle. (ii) There is no appearance of correlation coefficient $\rho$. Without the parameter $\rho$, we can avoid the error caused by the estimator of $\rho$. Moreover, we shall show in next section by simulation that the fitting effect by using the macroeconomic-risk model is also better than that by using Belkin et al's model.

Note that the assumption that the mean value of $Y$ is zero is not a strong limitation because one can reduce the general case to this special one through including a constant vector $\bm 1$ in the vector $\bm M$. This will cause no additional difficulty in the previous mathematical deduction.

\section{Simulation and Comparison}

In this section, we shall compare the effectiveness of one factor model with that of the macroeconomic-risk model in predicting \textbf{PD}.

The simulation is based on a used one factor model in practise, where the historical transition matrices are pulled out from S\&P Credit Pro, and the macroeconomic variables are chosen in the Federal Reserve supervisory scenarios. The series $Z_t$ has been extracted by following the work \cite{Bel-b}, and three macroeconomic variables has been selected to establish a linear regression model. According to \eqref{tres} and \eqref{est-rate}, a set of transition probability matrices $(P_{ij,t})$ can be derived.
We shall look on these transition probability matrices $(P_{ij,t})$ as the true transition probability matrices to compare the effectiveness of one factor model and our macroeconomic-risk model.  We use this treatment to get rid of the computational error caused by the null entries in the empirical transition probability matrices. In Table 1, a typical historical transition probability matrix is presented, whose many entries equal to zero, especially the transition probabilities from all grades to grade 8 are all equal to 0. This is obviously not reasonable.
\begin{center}
\begin{table}[!h]\label{c-1}
\begin{tabular}
[c]{|c|c|c|c|c|c|c|c|c|c|c|}%
\hline
1990Q1 & 1 & 2 & 3 & 4 & 5 & 6 & 7 &\ \ 8 \ \  & 9\\
\hline
1 & 99.65 & 0.35 & 0 & 0 & 0 & 0 & 0 & 0 & 0\\
2 & 0.37 & 96.33 & 2.21 & 0 & 0 & 1.10 & 0 & 0 & 0\\
3 & 0 & 0 & 96.15 & 0.77 & 1.54 & 1.54 & 0 & 0 & 0\\
4 & 0 & 0 & 0 & 94.90 & 1.02 & 3.06 & 0 & 0 & 1.02\\
5 & 0.99 & 0 & 0 & 1.48 & 92.11 & 3.94 & 0.98 & 0 & 0.50\\
6 & 0 & 0.87 & 0 & 0 & 1.74 & 94.78 & 0.87 & 0 & 1.74\\
7 & 0 & 0 & 0 & 0 & 0 & 2.78 & 83.33 & 0 & 13.89\\
8 & 0 & 0 & 0 & 0 & 0 & 0 & 0 & 0 & 0\\
\hline
\end{tabular}
\caption{Transition Matrix on the first quarter of 1990}
\end{table}
\end{center}
For computational concerning, the S\&P Credit rating grades were re-scaled into 9 grades in the following way:

{\small \makeatletter\def\@captype{table}\makeatother
\begin{tabular}
 [c]{|c|c|c||c|c|c|}%
 \hline
 No. & S\&\! P rating & model binning&No. & S\&\! P rating & model binning\\
 \hline
 1 & AAA & 1&12 & BB & 3\\
 2 & AA+ & 1&13 & BB- & 4\\
 3 & AA & 1&14 & B+ & 5\\
 4 & AA- & 1&15 & B & 6\\
 5 & A+ & 1&16 & B- & 6\\
 6 & A & 1&17 & CCC+ & 7\\
 7 & A- & 1&18 & CCC & 7\\
 8 & BBB+ & 2&19 & CCC- & 7\\
 9 & BBB & 2&20 & CC & 8\\
 10 & BBB- & 2&21 & C & 8\\
 11 & BB+ & 3&22 & D & 9\\
 \hline
 \end{tabular} }

\noindent The grade 9 denotes the default. Moreover, three selected macroeconomic variables are: CBOE Volatility Index for S\&P 500 Stock Price Index, Nominal GDP growth and Spread, Baa Corp Bond Yield less 10-Yr Treasury Yield.

The simulation and comparison is carried out in three steps.

\textbf{Step 1}:  Based on the priori extracted series $Z_t$, we add a normal perturbation to it, i.e.,
\[\widetilde Z_t=Z_t+\eta,\quad \text{where}\ \eta\ \text{is normally distributed random variable}.\] Inserting these $\widetilde Z_t$ into \eqref{est-rate}, we can get a set of transition probability matrices $(\widetilde P_{ij,t}^Z)$. These matrices $(\widetilde P_{ij,t}^Z)$ are viewed as the empirical sample of the true transition probability matrices $(P_{ij,t})$. Then, using the procedure of one factor model and using the selected three macroeconomic variables to establish the regression model, we can obtain the estimated transition probability matrices $(\hat P_{ij,t}^Z)$. The difference between $(P_{ij,t})$ and $(\hat P_{ij,t}^Z)$ reflects the accuracy of the estimation of one factor model.

\textbf{Step 2}: We still use the matrices $(\widetilde P_{ij,t}^Z)$ obtained in previous step as the empirical sample of $(P_{ij,t})$. Then, we can establish a regression model with respect to the three selected macroeconomic variables. Hence, we get another set of estimation $(\hat P_{ij,t}^{New})$ of $(P_{ij,t})$ via \eqref{e-6}-\eqref{e-7}, which is the estimation provided by the macroeconomic-risk model.

\textbf{Step 3}: For every estimated transition probability matrix, the probability of default of each grade can be easily calculated.  For example, $P_{k9}$ is the probability of default for an individual with initial  rating grade $k$, which is denoted by $\mathbf{PD}_k=P_{k9}$.

\begin{figure}[!h]\centering
  \includegraphics[width=0.8\textwidth]{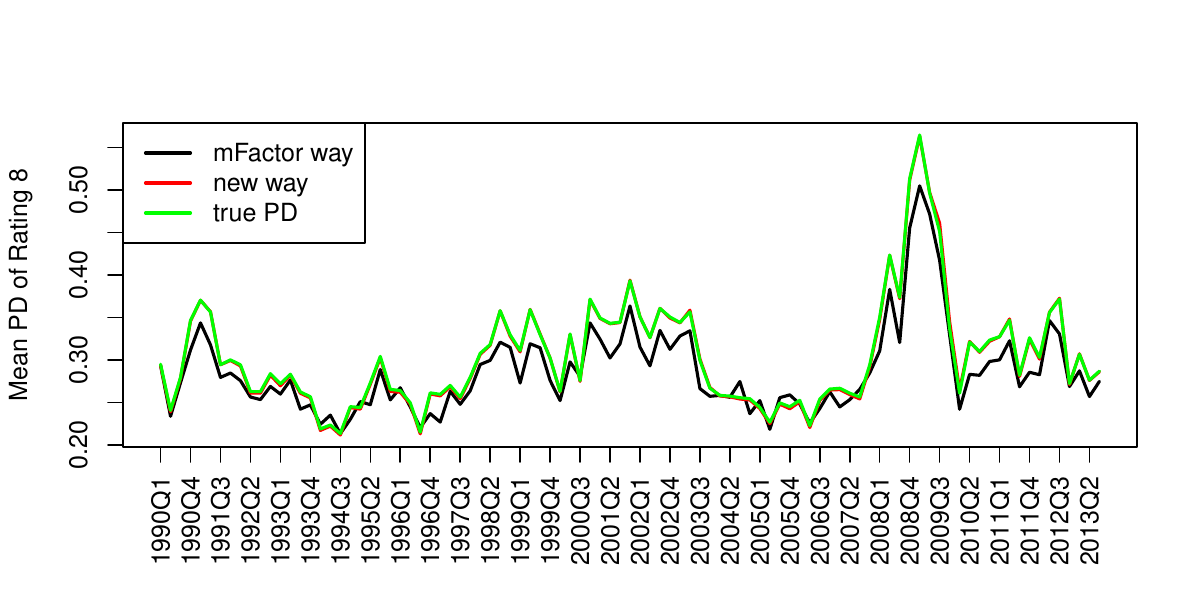}\\
  \caption{$PD_8$}\label{fig1}
\end{figure}

As an example, we present the comparison result of $\textbf{PD}_8$ in Figure \ref{fig1}. In Figure \ref{fig1}, green line corresponds to the true value of $P_{89,t}$; red line shows the estimated $\hat P_{89,t}^{New}$ of macroeconomic-risk model; black line is associated with the estimated $\hat P_{89,t}^Z$. Figure \ref{fig1} shows that the red line and the green line  almost stick together, however, the black line is significantly diverse from the green line. Actually, $\textbf{PD}_k$ for $k=1,\ldots,7$ all present such trait (see Figure \ref{fig4} in Appendix).

Although the macroeconomic-risk model presented an outstanding performance in previous test, one may doubt the consistency of such good performance. Next, we repeat previous three steps for 1000 times to check this model's performance. For every time, we can get two series of estimated transition probability matrices $(\hat P_{ij,t}^Z)$ and $(\hat P_{ij,t}^{New})$. Then, we compare the mean-square error (MSE) of these two methods. The result for $\textbf{PD}_8$ is presented in Figure 2, which clearly shows that our method is better than that of one factor model. Other graphs for $\textbf{PD}_{k}$ for $k=1,\ldots,7$ are given in Figure 3 in Appendix.

\begin{figure}[!h]\centering
  \includegraphics[width=0.8\textwidth]{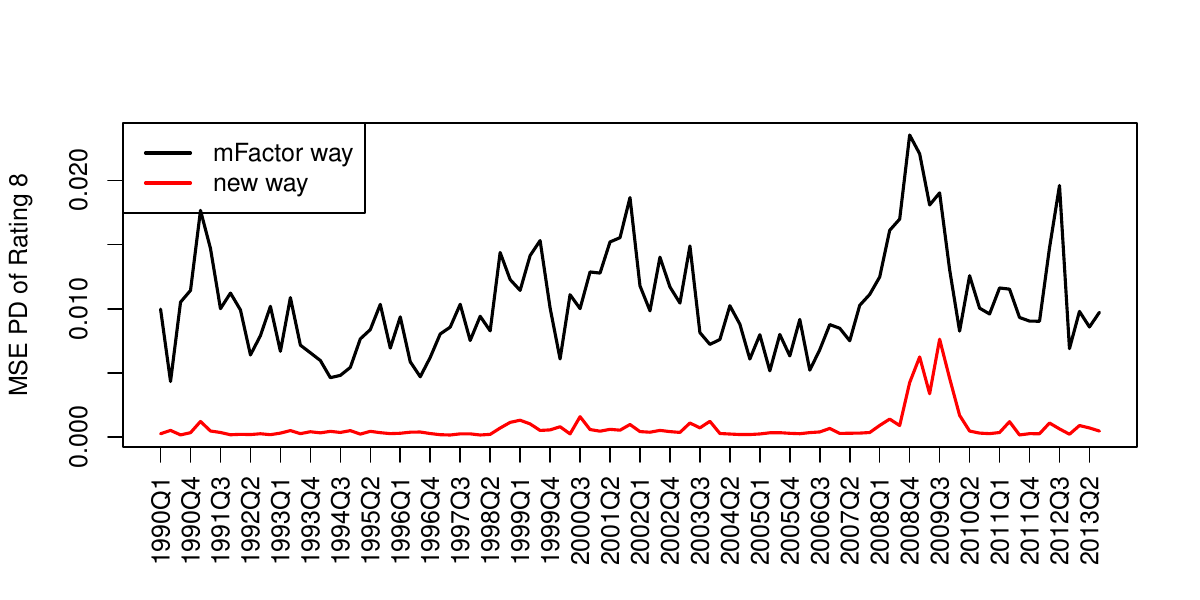}\\
  \caption{MSE of $PD_8$}\label{fig2}
\end{figure}
In Figure \ref{fig2}, black line is the MSE curve for one factor model, red line is MSE curve for macroeconomic-risk model.
This graph clearly shows that, in $1000$ times average,
the estimated \textbf{PD} of macroeconomic-risk model is still closer to its true value.

Consequently, these simulation results demonstrate that the macroeconomic-risk model could provide a more effective estimate of \textbf{PD} than one factor model.

\noindent{\bf Appendix}.

\begin{figure}[!h]\centering
\subfigure[MSE OF $PD_1$]{
  \includegraphics[width=0.45\textwidth]{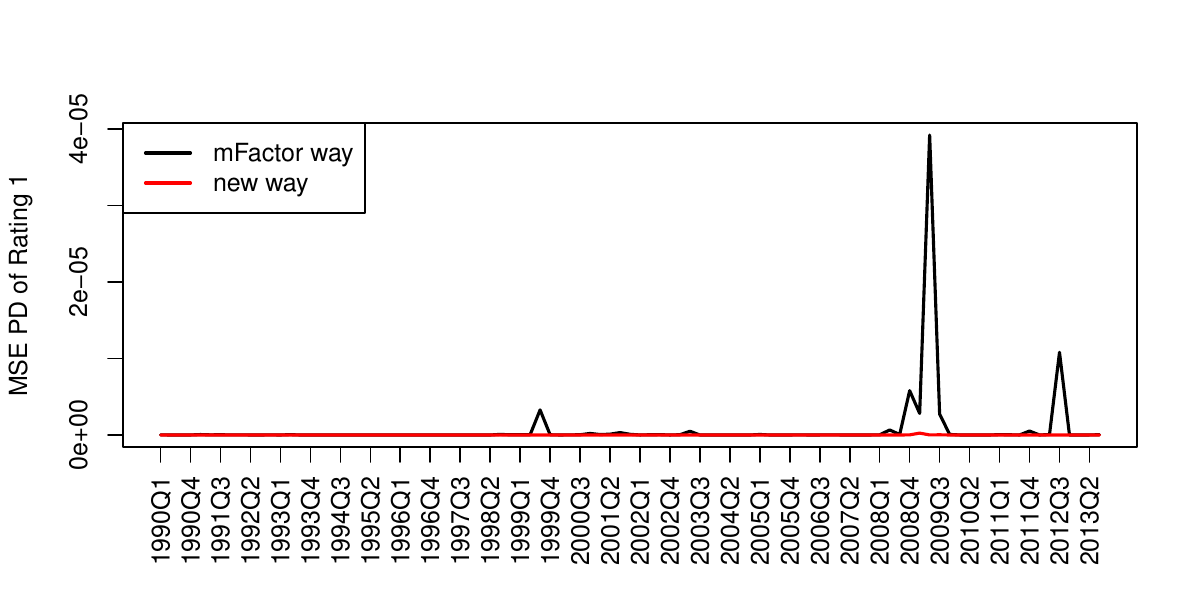}}
\subfigure[MSE OF $PD_2$]{
  \includegraphics[width=0.45\textwidth]{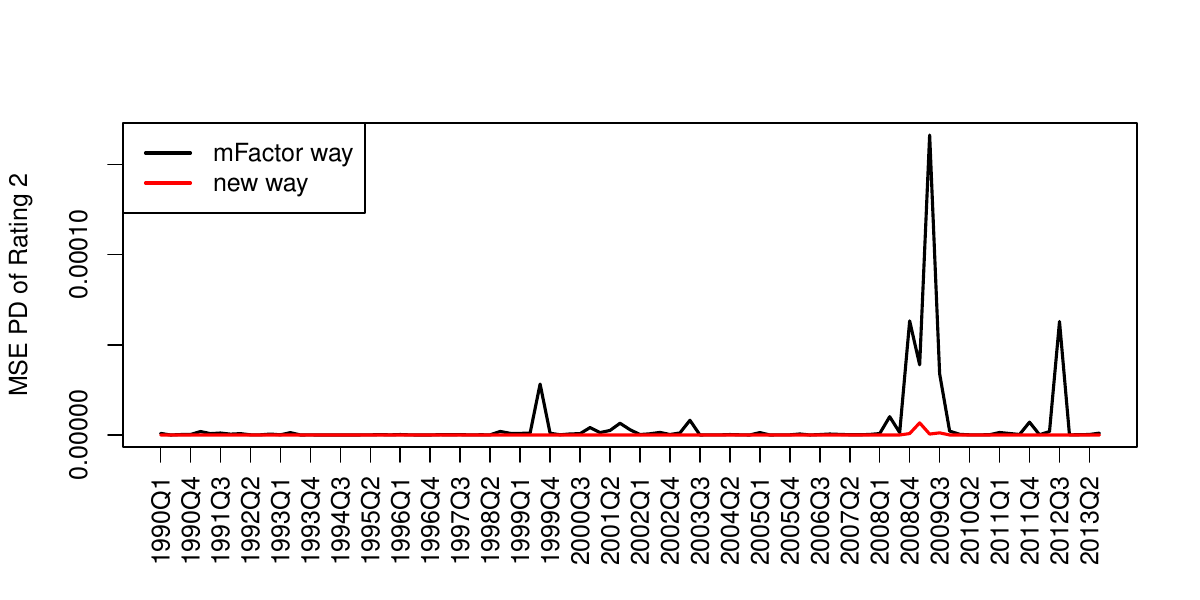}}\\
\subfigure[MSE OF $PD_3$]{
  \includegraphics[width=0.45\textwidth]{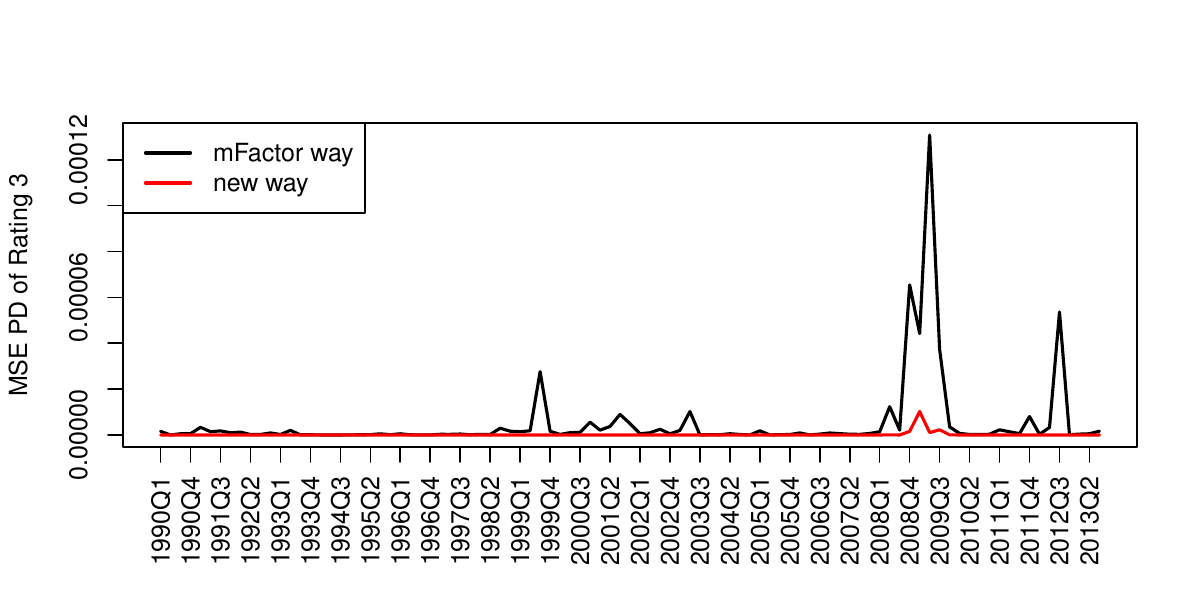}}
\subfigure[MSE OF $PD_4$]{
  \includegraphics[width=0.45\textwidth]{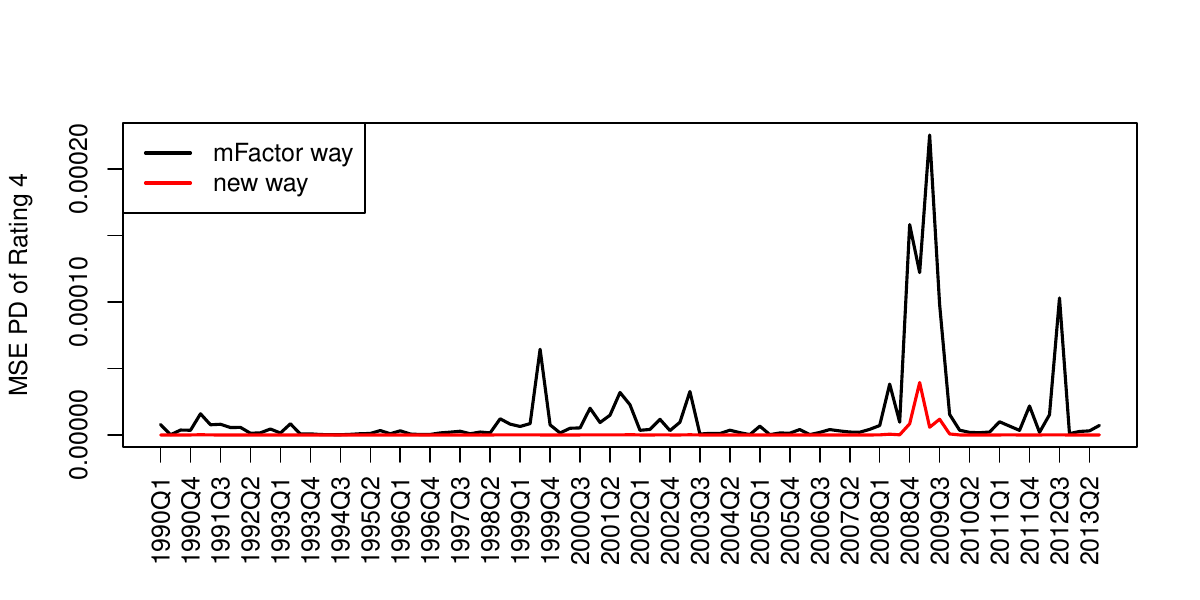}}\\
\subfigure[MSE OF $PD_5$]{
  \includegraphics[width=0.45\textwidth]{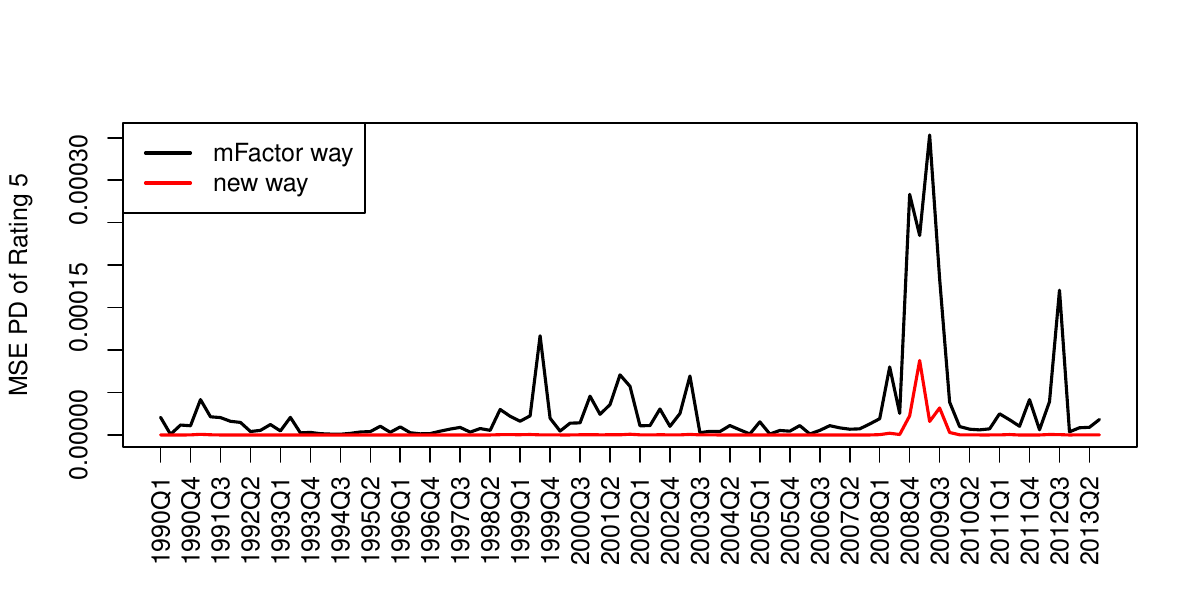}}
\subfigure[MSE OF $PD_6$]{
  \includegraphics[width=0.45\textwidth]{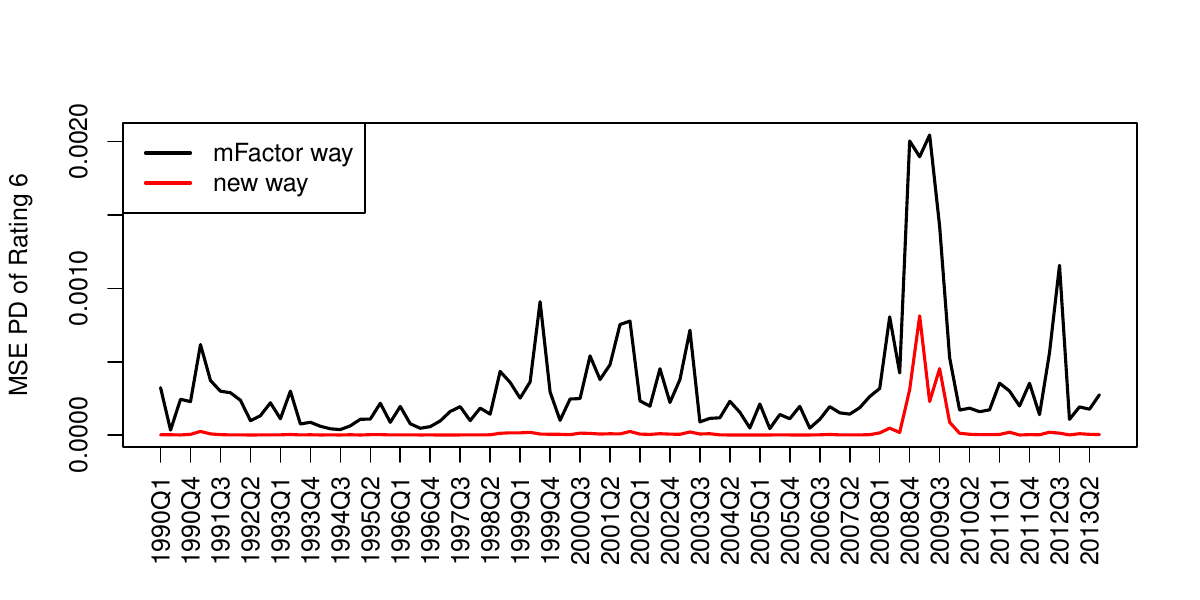}}\\
\subfigure[MSE OF $PD_7$]{
  \includegraphics[width=0.45\textwidth]{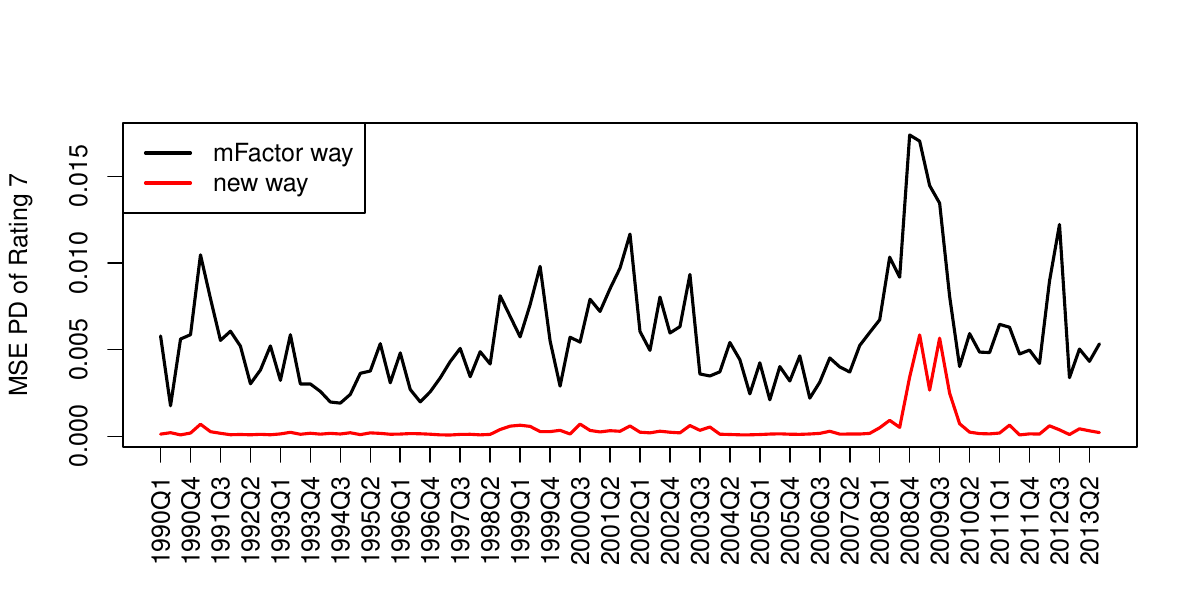}}
\subfigure[MSE OF $PD_8$]{
  \includegraphics[width=0.45\textwidth]{MSEPD8.pdf}}
  \caption{MSE OF $PD$}
\end{figure}

\begin{figure}[!h]\centering \label{fig4}
\subfigure[$PD_1$]{
  \includegraphics[width=0.45\textwidth]{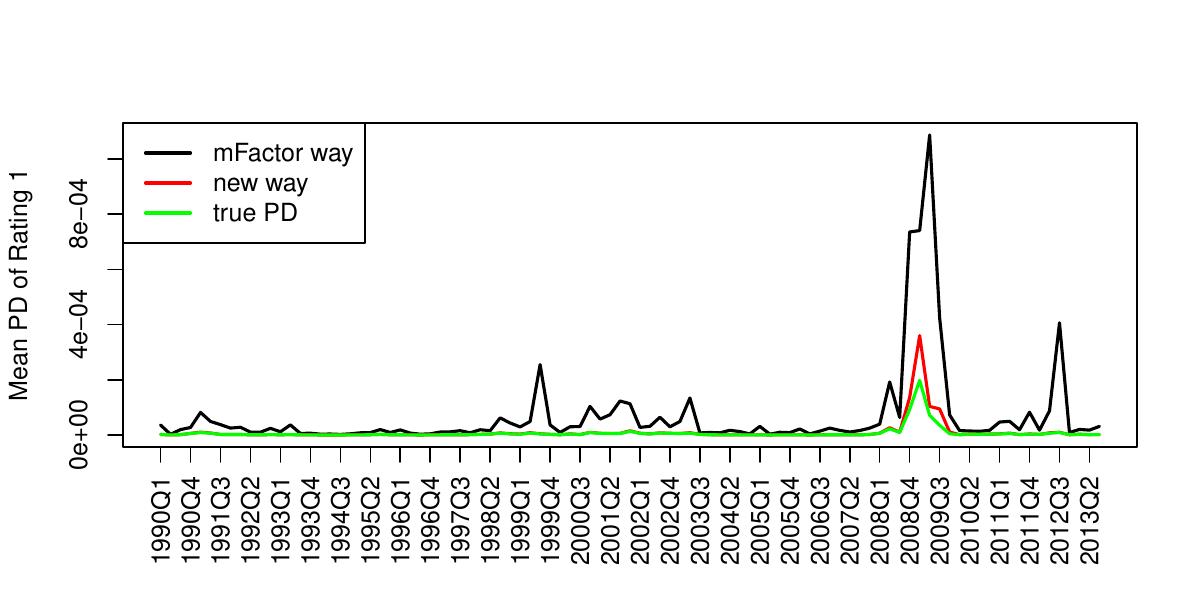}}
\subfigure[$PD_2$]{
  \includegraphics[width=0.45\textwidth]{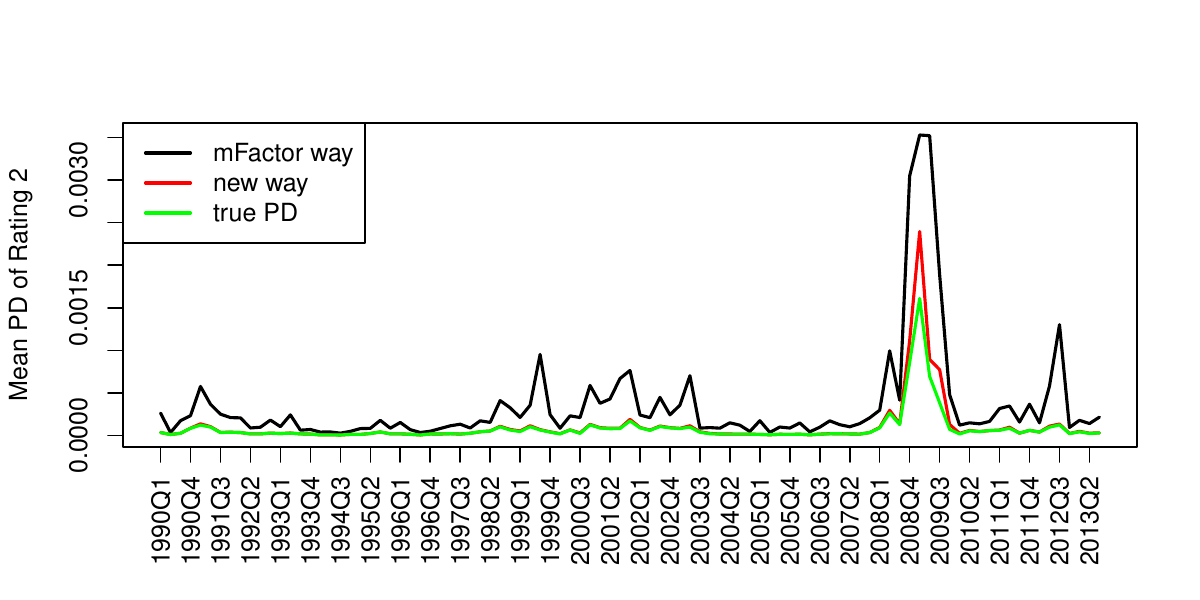}}\\
\subfigure[$PD_3$]{
  \includegraphics[width=0.45\textwidth]{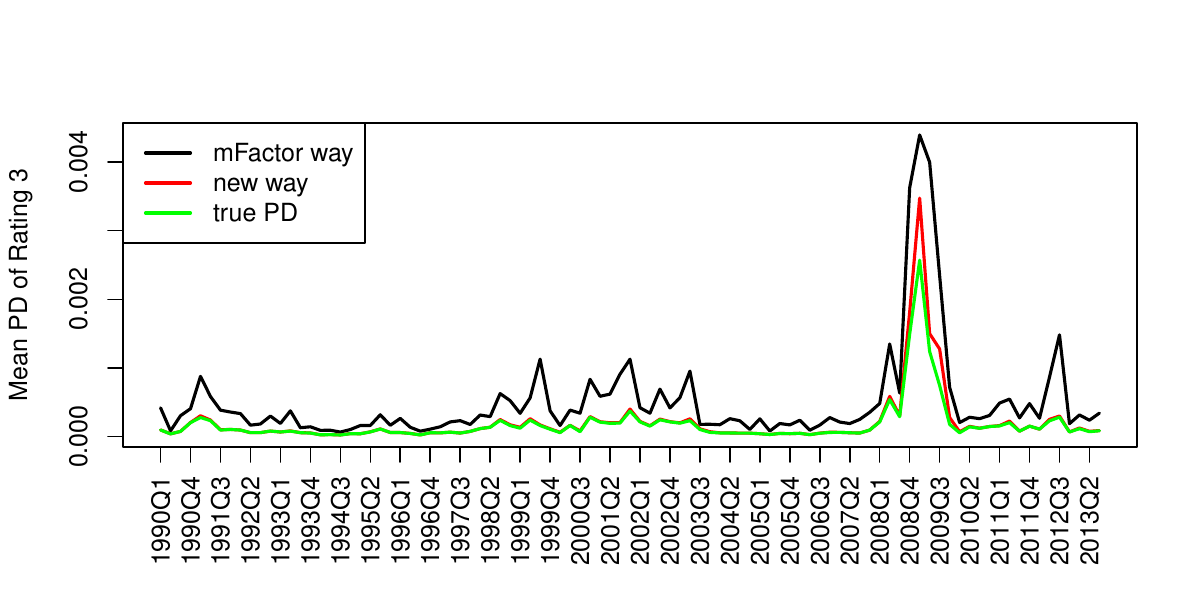}}
\subfigure[$PD_4$]{
  \includegraphics[width=0.45\textwidth]{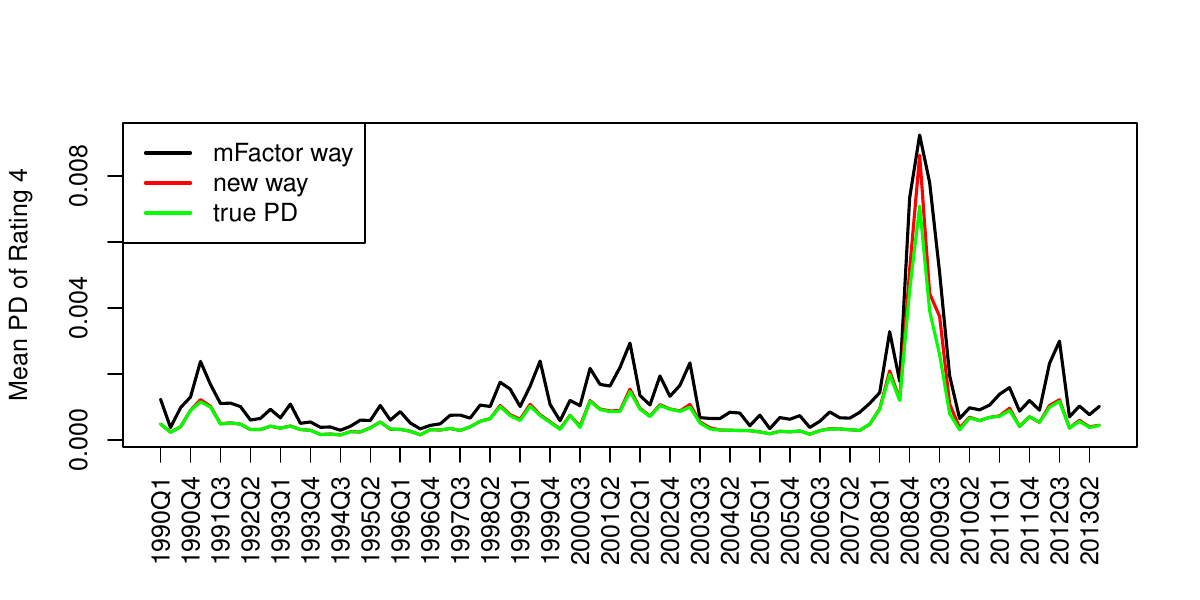}}\\
\subfigure[$PD_5$]{
  \includegraphics[width=0.45\textwidth]{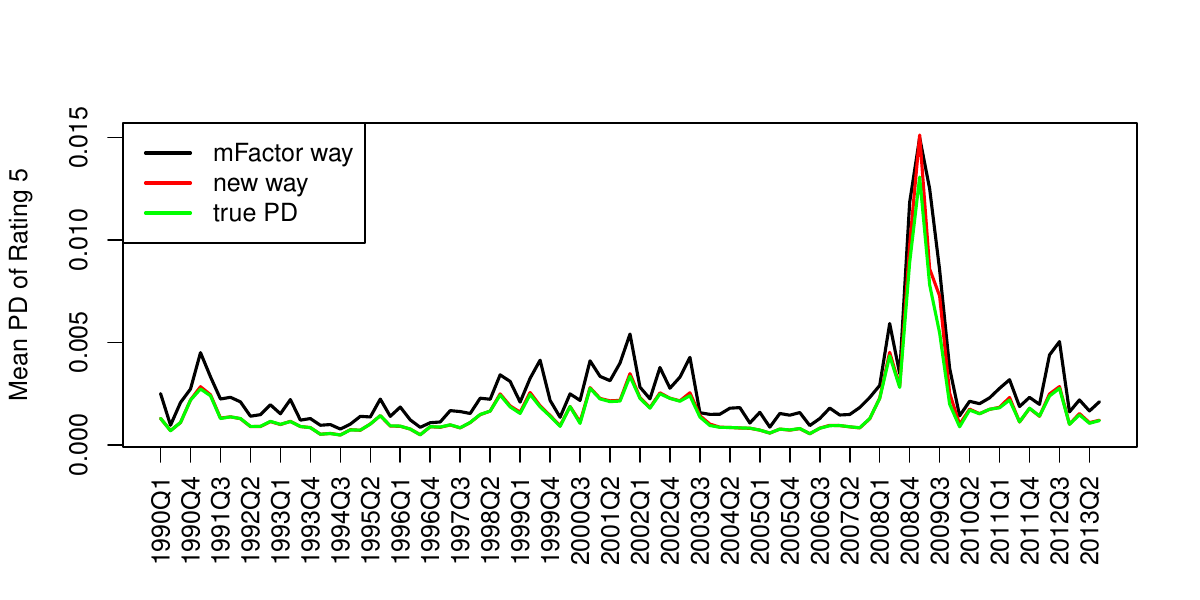}}
\subfigure[$PD_6$]{
  \includegraphics[width=0.45\textwidth]{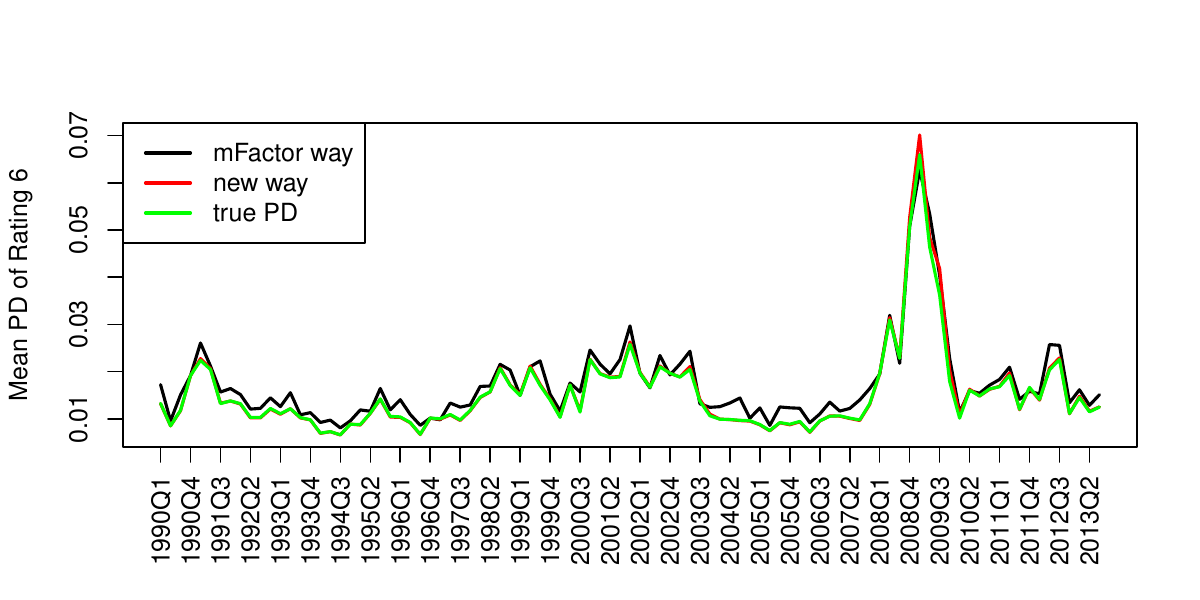}}\\
\subfigure[$PD_7$]{
  \includegraphics[width=0.45\textwidth]{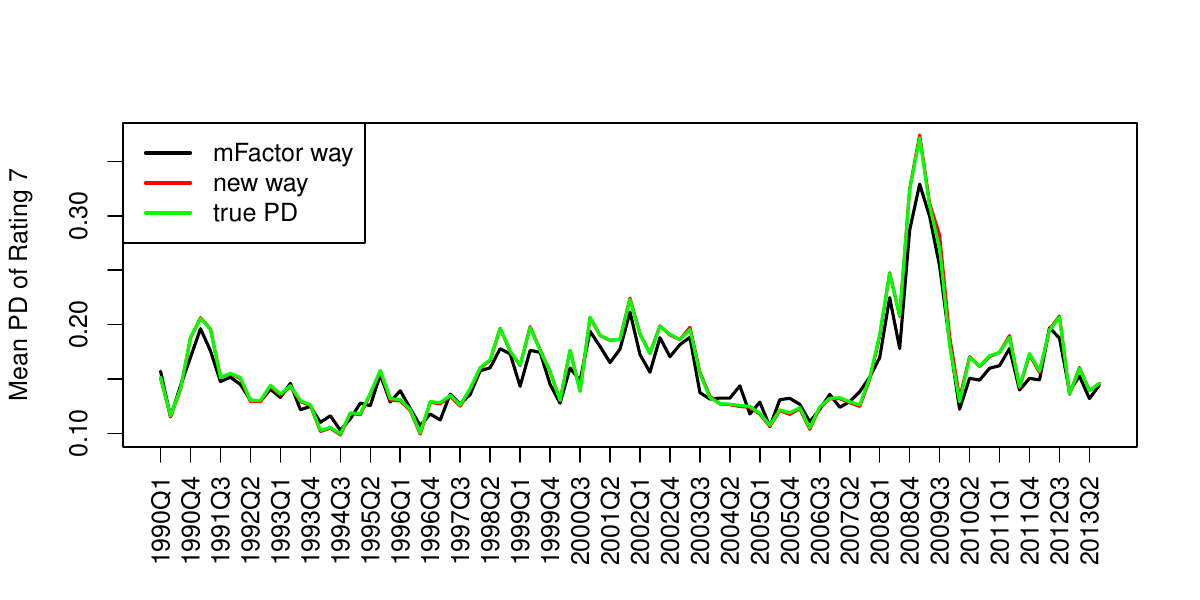}}
\subfigure[$PD_8$]{
  \includegraphics[width=0.45\textwidth]{PD8.pdf}}
  \caption{One time simulation $PD$}
\end{figure}

\end{document}